\documentclass[sigconf]{nimeart}
\usepackage{tabularx}
\usepackage{booktabs}
\usepackage{tablefootnote}
\usepackage[utf8]{inputenc}
\usepackage{fontspec}

\AtBeginDocument{%
  }

\setcopyright{cc}
\copyrightyear{2026}
\acmYear{2026}
\acmDOI{}

\acmConference[NIME '26]{International Conference on New Interfaces for Musical Expression}{June 23--26,
  2026}{London, United Kingdom}
\acmISBN{}

\begin{document}

\title{RenCon 2025: Revival of the Expressive Performance Rendering Competition}

\author{Huan Zhang}
\email{huan.zhang@qmul.ac.uk}
\orcid{0000-0000-0000-0000}
\affiliation{%
  \institution{Queen Mary University of London}
  \city{London}
  \country{UK}
}

\author{Taegyun Kwon}
\email{ilcobo2@kaist.ac.kr}
\affiliation{%
  \institution{Korea Advanced Institute of Science and Technology}
  \city{Daejeon}
  \country{Korea}
}

\author{Anders Friberg}
\email{afriberg@kth.se}
\affiliation{%
  \institution{KTH Royal Institute of Technology}
  \city{Stockholm}
  \country{Sweden}
}

\author{Junyan Jiang}
\email{jj2731@nyu.edu}
\affiliation{%
  \institution{New York University}
  \city{New York}
  \state{New York}
  \country{USA}
}

\author{Hayeon Bang}
\email{hayeonbang@kaist.ac.kr}
\affiliation{%
  \institution{Korea Advanced Institute for Science and Technology (KAIST)}
  \city{Daejeon}
  \country{South Korea}
}

\author{Hyeyoon Cho}
\email{hyeyooncho@kaist.ac.kr}
\affiliation{%
  \institution{Korea Advanced Institute for Science and Technology (KAIST)}
  \city{Daejeon}
  \country{South Korea}
}

\author{Gus Xia}
\email{gus.xia@mbzuai.ac.ae}
\affiliation{%
  \institution{Mohamed bin Zayed University of Artificial Intelligence}
  \city{Abu Dhabi}
  \country{UAE}
}

\author{Akira Maezawa}
\email{akira.maezawa@music.yamaha.com}
\affiliation{%
  \institution{Yamaha Corporation}
  \city{Hamamatsu}
  \country{Japan}
}

\author{Simon Dixon}
\email{s.e.dixon@qmul.ac.uk}
\affiliation{%
  \institution{Queen Mary University of London}
  \city{London}
  \country{UK}
}

\author{Dasaem Jeong}
\email{dasaem.jeong@gmail.com}
\affiliation{%
  \institution{Sogang University}
  \city{Seoul}
  \country{South Korea}
}

\renewcommand{\shortauthors}{Zhang et al.}

\begin{abstract}
This paper presents a comprehensive documentation of RenCon 2025, the revival of the expressive performance rendering competition which took place at ISMIR 2025 in Daejeon, Korea. The competition attracted 9 entries from international research groups, representing diverse approaches to expressive piano performance rendering. The two-phase assessment structure comprised a preliminary online evaluation and live real-time rendering at the conference. We analyze the competition format, participant demographics, system performance, and lessons learned for future iterations. The results demonstrate significant advances in expressive rendering capabilities while highlighting remaining challenges in achieving human-level musical expression.
\end{abstract}


\maketitle

\section{Introduction}

Expressive performance rendering—the task of transforming a musical score into a performance with human-like timing, dynamics, and articulation, has been a central challenge in computer music since the 1980s \cite{Sundberg2010AttemptsRules}. The RenCon (CONtest for performance RENdering systems) series, initiated in 2002, provided an interactive evaluation platform for comparing different approaches to this challenge \cite{Hiraga2002RenconSystems}. 

After the last competition in 2013, RenCon disappeared from the research landscape, coinciding with rapid advances in machine learning and artificial intelligence. Along with the creation of large-scale performance datasets such as ATEPP \cite{Zhang2022ATEPPPerformance} and ASAP \cite{Peter2023AutomaticDataset}, the field has since witnessed the emergence of neural approaches to music generation and performance modeling, making the revival of systematic evaluation more important than ever.
RenCon 2025 marked the return of this competition, hosted as part of the Music Information Retrieval Evaluation eXchange (MIREX) at the International Society for Music Information Retrieval (ISMIR) Conference in 2025. This paper presents a comprehensive documentation of the competition design, participating systems, evaluation methodology, and results, contributing insights for future performance rendering evaluation and the broader Music Information Retrieval (MIR) community \footnote{Competition website: \url{https://ren-con2025.vercel.app/} \\ Online audition: \url{https://ren-con2025-audition-page.vercel.app/} \\ Results: \url{https://github.com/ismir-mirex/RenCon2025}}. 

\section{Past Competitions}

The Contest for Performance Rendering Systems (RenCon) was established to encourage research in performance rendering, and address the lack of standardized evaluation methods for computer systems for expressive music performance. Initiated in 2002, RenCon served as a ``Turing Test" for musical expression, providing a public forum where autonomous and human-supported rendering systems could be judged by both experts and general audiences \cite{Hiraga2002RenconSystems}.

In its early iterations, the competition focused on defining valid evaluation criteria for the subjective nature of musical performance. \citet{Hiraga2004RenconExpression} introduced specific methodologies such as the ``Turing Test" and ``Gnirut Test" (a reverse Turing Test) to assess whether a system's output could be distinguished from human performance. These contests highlighted the challenge of bridging the gap between technical system measurements and human aesthetic perception.

By 2008, RenCon had matured into a multi-category event, distinguishing between fully autonomous systems and those requiring human intervention. \citet{Hashida2008RenconSystems} detailed the 2008 contest at the ICMPC10 conference, noting the success of systems in the autonomous section. This period solidified RenCon's role not just as a competition, but as a research platform for analysing the relationship between algorithmic parameters and listener preference.

The cumulative knowledge of the competition's decade-long run was synthesized by Katayose et al. \cite{Katayose2012OnExperience}. This retrospective analysis emphasized the ``RenCon Experience," discussing the difficulty of maintaining a ``common ground" for evaluation as systems became more diverse. Despite the competition's discontinuation after 2013, its legacy remains critical.

For an overview of the different RenCon workshops, see Table~\ref{tab:overview}. Throughout the years there has been some experimentation of the contest procedure regarding evaluation, preparation of performances, and selection of music. However, the compulsory music has been quite restricted mostly to Chopin and Mozart and has always been restricted to the Western classical repertoire. The last workshop in Stockholm 2013 was organized together with the SMAC/SMC conference using a Disklavier piano. Two pieces were used, one by Domenico Scarlatti and one by Nino Rota. The pieces were performed live and votes were collected from both a jury and the audience. The winner in the automatic category was the Stochastic model \cite{okumura2011stochastic} and in the human-in-the-loop category it was VirtualPhilharmony \cite{baba2010virtualphilharmony} .

\begin{table*}[h]
\centering
\caption{Overview of past RenCon workshops.}
\label{tab:overview}
\resizebox{\textwidth}{!}{%
\begin{tabular}{lllll}
\hline
\textbf{Year} & \textbf{Conference}    & \textbf{Piano}     & \textbf{Music}  & \textbf{Interaction} \\
\hline
2002 & ICAD, Kyoto         & Disklavier                & Open                                         & Manual; Autonomous              \\
2002 & FIT, Tokyo          & MIDI Bar on piano & Compulsory (Mozart or Chopin piece) & Manual; Assisted; Autonomous \\
2003 & IJCAI, Acapulco     & GigaPiano                 & Compulsory (sel. Chopin)$+$Open                & -                              \\
2004 & NIME, Hamamatsu     & GigaPiano                 & Compulsory (any Chopin or Mozart)$+$Open       & -                              \\
2005 & ICMC, Barcelona     & GigaPiano                 & Compulsory (Mozart)$+$Open                     & -                              \\
2006 & ISMIR, Victoria     & Bösendorfer               & Compulsory (Chopin)$+$Open                     & -                              \\
2008 & ICMPC, Sapporo      & Disklavier                & One or two pieces with limited time          & Assisted; Autonomous \\
2009 & EC, Tokyo           & Not known                 & Not known                                    & Assisted; Autonomous \\
2011 & SMC, Padova         & Disklavier                & New  piece$+$a selected piece                  & Assisted;  Autonomous \\
2013 & SMAC/SMC, Stockholm & Disklavier                & Compulsory (Rota and Scarlatti)              & Assisted; Autonomous \\
\hline
\end{tabular}
}
\end{table*}

Problems due to the differences in dynamics across different digital pianos and Disklaviers have been an issue since the start of RenCon. The reason is that MIDI velocity is not related to any measurable quantities such as sound level. Instead, the MIDI velocity response has been selected ad-hoc and is different for different synthesizers, MIDI keyboards, and piano simulations. Since the different databases of piano performances that exist usually are coded in MIDI velocity as given by the instrument, the lack of dynamics calibration is a general problem in the research community. The long-term solution would be to translate the MIDI velocity to sound level for the instruments used for data collection  \cite{bresin2002director}.

This problem of MIDI velocity was addressed already in the 2003 workshop in which the commercial sampling synthesizer GigaPiano was used, thus the piano sampling could be used by the participants in advance. However, that lacked the realism of a real acoustic piano, and subsequently the later RenCons returned to the Disklavier or used the Bösendorfer SE system. A solution specifically for the RenCon workshop would be to provide a simulation of the instrument that is going to be used. In the case of the Disklavier, it would be possible, for example, to sample the instrument in the workshop using an automatic procedure and provide it as a sampling library. In this way, developers could test their models in advance using proper velocity mapping. Another option would be to have access to the instrument in advance so that it is possible to make adjustments before the actual workshop. This has the disadvantage that it would be rather limited in time and that only one user can test it at a time.

\section{Competition Design}

RenCon 2025 adopted a two-phase structure consisting of an online audition round and a live final round. The online audition enabled broad international participation by removing the need for travel and allowing evaluators and teams to engage asynchronously across time zones. Based on the audition outcomes, the live final round focused on a smaller number of shortlisted teams, ensuring higher overall quality while keeping the session feasible within the limited time available during the conference.

\subsection{Phase 1: Preliminary Round (Online)}
\begin{itemize}
\item Submission: May 30 - August 25, 2025, through MIREX's online portal 
\item Task: Render 3 pieces (2 required + 1 free choice)
\item Format: MusicXML input, MIDI or WAV output
\item Evaluation: Online voting
\item Duration limit: 5 minutes total for all pieces
\end{itemize}

The preliminary round featured four designated pieces in MusicXML format, from which the entrants chose two:
\begin{itemize}
\item Handel: Capriccio in G minor, HWV 483
\item Beethoven: 32 Variations in C minor, WoO 80 - Theme and the first 5 variations
\item Rachmaninoff: Здесь хорошо (How Fair this Spot), Op. 21, No. 7 (transcribed for solo piano \footnote{We obtained the arranged version from Musescore: \url{https://musescore.com/user/33600951/scores/10812568}, and we obtained the arranger's permission to use it in this competition.})
\item Amy Beach: Eskimos, Op.64, No.4 - With Dog-Teams
\end{itemize}

To ensure fair and transparent evaluation, all post-processing applied to preliminary submissions was documented in the submission reports. In particular, symbolic (MIDI) outputs were rendered to audio using a consistent physical instrument setup: a Disklavier in the Vienna office of the Johannes Kepler University Linz. Participants were required to disclose any additional processing, human intervention, or manual editing (e.g., MIDI cleanup), and submissions were expected to minimise such interventions unless explicitly justified. However, we also allowed audio-based submissions, given the recent surge in audio generative modeling approaches \cite{zhang2025renderbox, jung2024continu}.

\begin{figure}
    \centering
    \includegraphics[width=\linewidth]{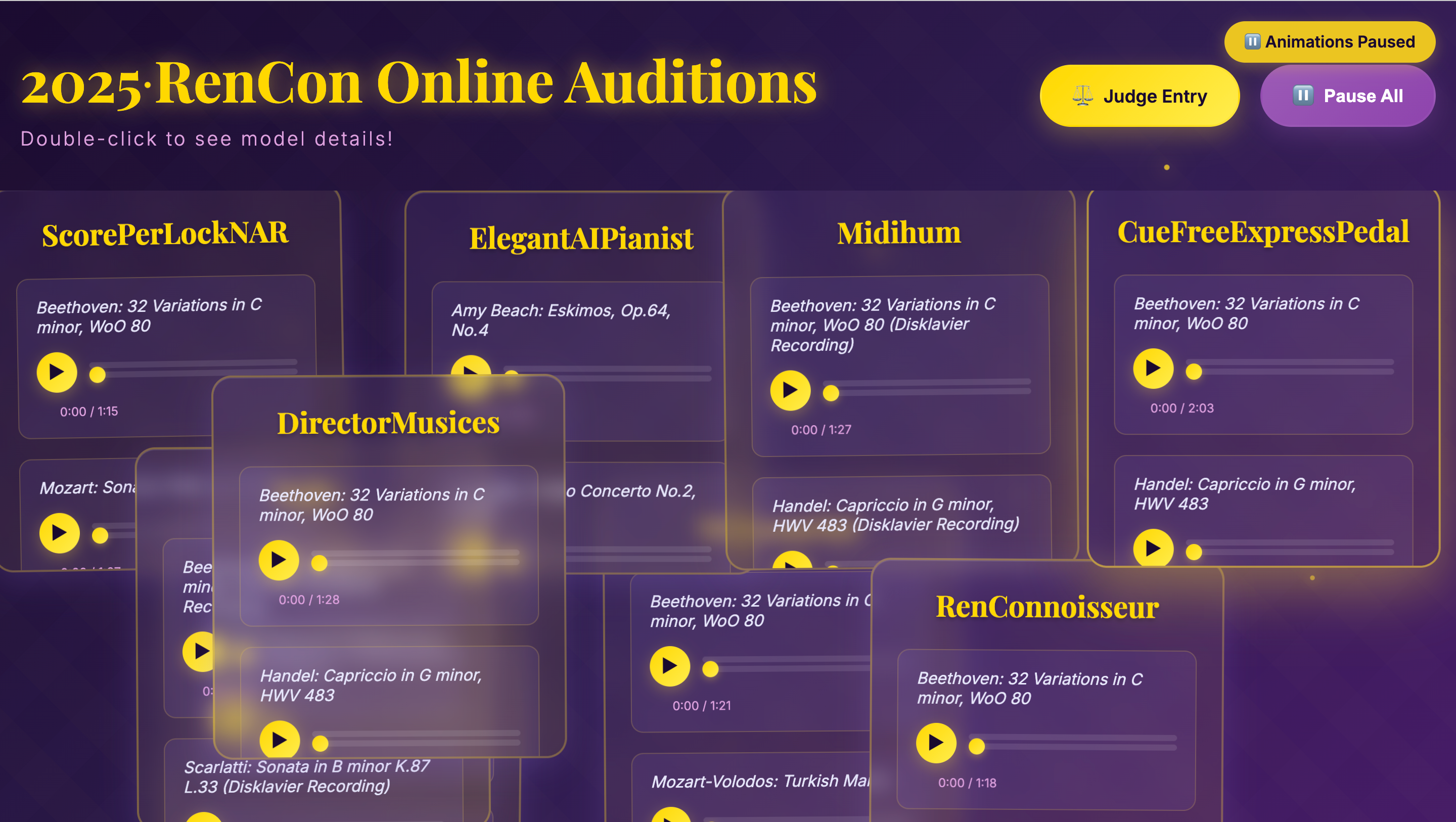}
    \caption{Online audition platform}
    \label{fig:online_audition}
\end{figure}

Figure~\ref{fig:online_audition} shows the website of the online audition submissions. Each submission occupies a floating block, which can be double-clicked to pop up. A brief summary of the model's technique is provided, with the audio playback in the block. On the top right of the website, the listener can submit their evaluations through a Google form. 
During the evaluation, the names of the teams or models were anonymised. The page was then de-anonymised after the online results were released.

\subsection{Phase 2: Live Contest}
\begin{itemize}
\item Date: September 25, 2025 (evening of ISMIR final day)
\item Task: Rendering of an unseen piece under a time constraint (48 hours)
\item Evaluation: Live audience voting at concert
\end{itemize}

\begin{figure}
    \centering
    \includegraphics[width=\linewidth]{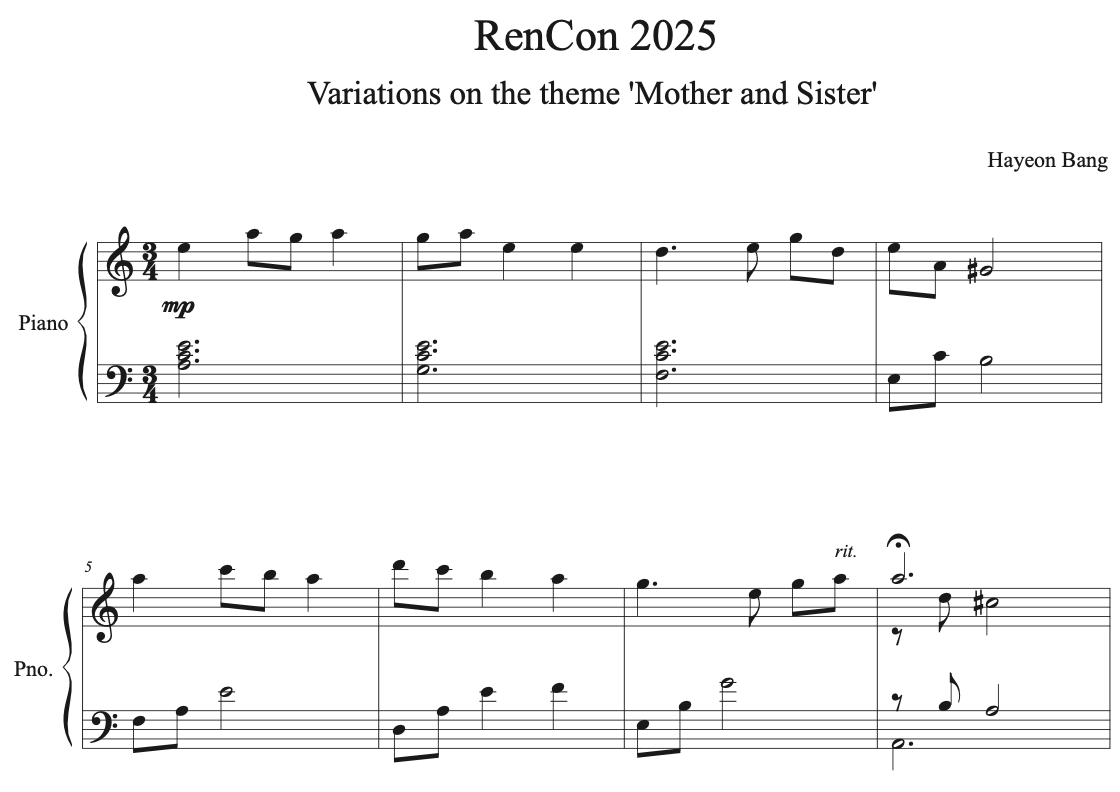}
    \caption{Theme of the final piece. The full score is available in the github repository. }
    \label{fig:final_piece_theme}
\end{figure}

The live contest's unseen piece was a new composition by Hayeon Bang, distributed to the participants on September 23, 2025. The theme, shown in Figure~\ref{fig:final_piece_theme}, is an eight-measure melody adapted from the Korean folk song 'Eommaya Nunaya (Mother and Sister)'. The piece follows a theme and variation form, comprising 94 measures and four variations.
Each variation was intentionally composed to imitate the musical styles of Bach, Mozart, Chopin, and Rachmaninoff, respectively, in order to evaluate the participants' ability to perform diverse stylistic traditions.

\textbf{Human baseline}: To make the competition a ``music performance Turing test", we invited a pianist to serve as the human baseline. Hyeyoon Cho studied piano performance at the University of Texas and earned both a Master of Music and a Performer’s Diploma at Indiana University. A co-winner of the Nilsson Piano Competition, she has performed as a soloist with the Indiana University Student Orchestra and served as an instructor for the Young Pianists Program.
She spent a total of seven hours over five days practising the final round piece. 
Her performance was recorded on the Disklavier on the day of the live competition, and then played back in the concert among the submissions. The submissions were played anonymously, in random order. 
We used the same evaluation form as the online round, with one additional question for the audience to identify the human performance. 

\subsubsection{Venue Details}

The KAIST Auditorium serves as a mid-sized performance space for cultural and academic events on campus. Containing around 500 seats, the hall is regularly used for piano recitals, ensemble performances, and ticketed concerts. Its layout and capacity provide an intimate yet formal listening environment suitable for both acoustic and amplified performances.

\subsubsection{Disklavier Calibration}

Prior to the concert, organizers performed one round of calibration on the Disklavier, including team-specific volume balancing and pedal response checks, to adapt to the hall acoustics and velocity level.

\begin{figure}
    \centering
    \includegraphics[width=0.5\linewidth]{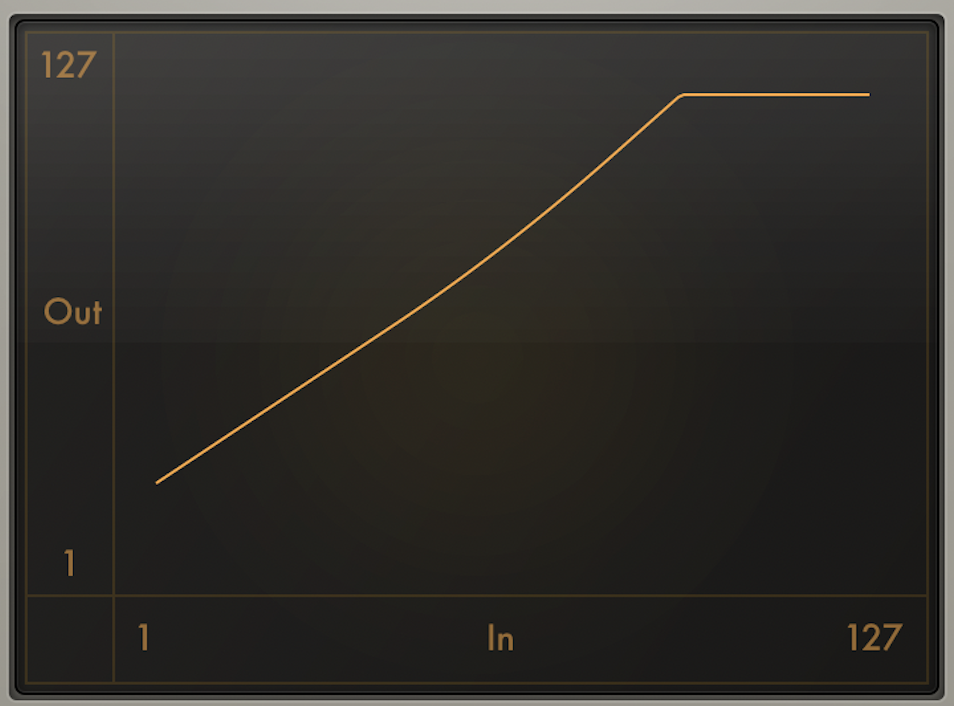}
    \caption{Logic Pro MIDI Velocity Processor setting used to raise low-velocity playback (example from adjustment for YQX+).}
    \label{fig:midi-velocity-processor}
\end{figure}

Because most teams could not directly audition and adjust their submitted MIDI files against the exact venue instrument in advance, we used a heuristic global velocity remapping profile while listening to the playback in Logic Pro. We used the MIDI Velocity Processor plug-in to adjust the offset and slope (Figure~\ref{fig:midi-velocity-processor}).
Pedaling information also varied across submissions. Since performances with no predicted sustain pedal were easily distinguishable in listening, we applied a conservative half-pedal setting only to MIDI files without pedal prediction.

\begin{figure}
  \centering
  \includegraphics[width=\linewidth]{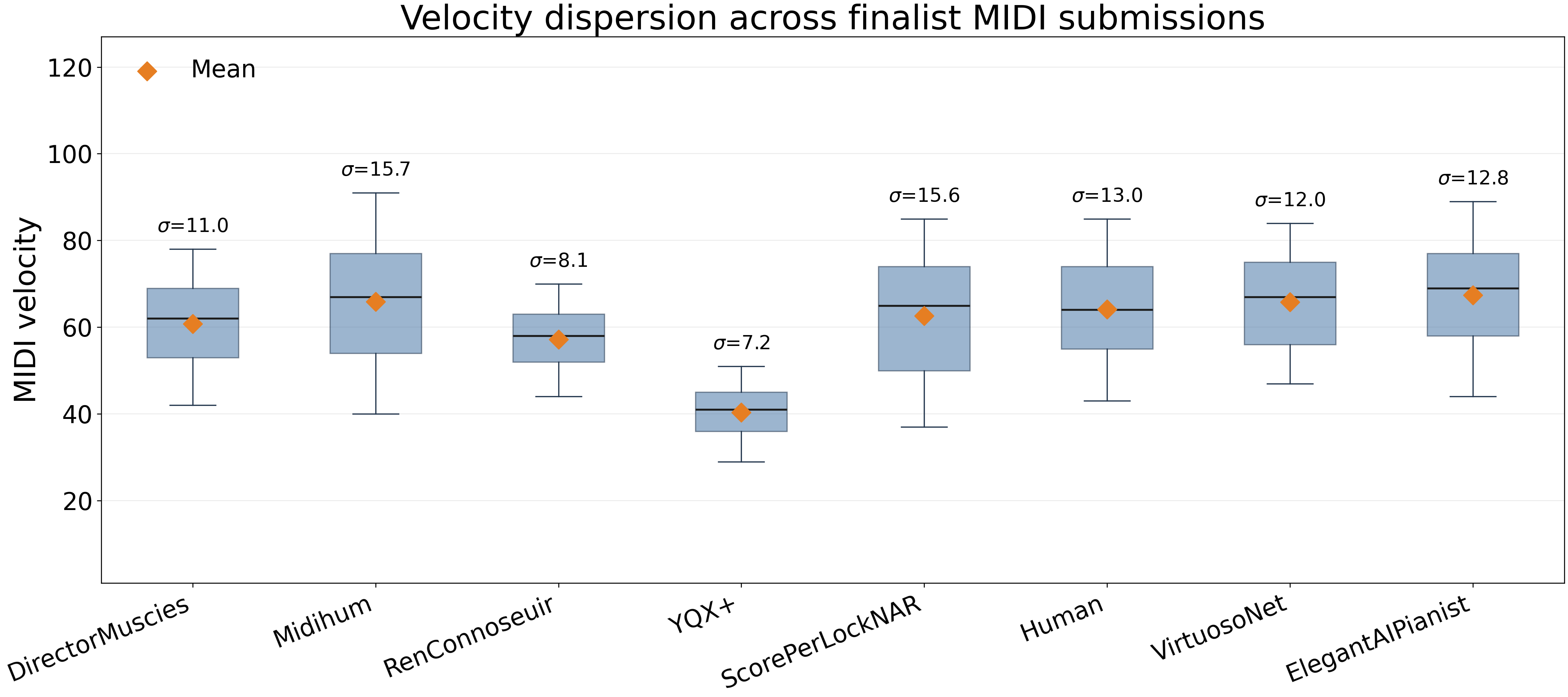}
  \caption{Velocity box plot for live-contest MIDI submissions. Boxes show interquartile ranges, whiskers indicate the 5th--95th percentiles, orange diamonds indicate means, and text annotations show standard deviation $\sigma$.}
  \label{fig:live-velocity-box}
\end{figure}

Figure~\ref{fig:live-velocity-box} summarizes velocity dispersion from one representative MIDI file per finalist team (excluding only the audio-only Contin-U). The spread differs substantially across systems, which supports the need for a calibration step.

\section{Evaluation}

The evaluation was conducted using a structured online Google form designed to collect both quantitative ratings and qualitative feedback. For each team's rendering, the evaluator is asked to rate it on a 5-point Likert scale: \textit{How would you rate [model]'s performance?} This accounts for the overall perception (naturalness, expressivity, suitability, etc.) of the audience. The evaluator is also asked for their global confidence score from 1 to 5, which is used to weight their ratings. 

In addition to score-based judgments, evaluators were invited to provide open-ended comments on model quality and the overall competition. Optional demographic questions (musical background and listening habits) were included to contextualise the responses.

\subsection{Preliminary Round Evaluation}

The online evaluation used a weighted voting system, where participants self-rated their expertise on a 1-5 star scale and their scores were then weighted accordingly.

\begin{table}[h]
\centering
\caption{Evaluator demographics for preliminary and final rounds (multiple selections allowed).}
\label{tab:demographics}
\begin{tabular}{lcc}
\hline
\textbf{Category} & \textbf{Preliminary (N=24)} & \textbf{Final (N=48)} \\
\hline
Researchers            & 12 (54.5\%) & 33 (71.7\%) \\
Music technologists     & 10 (45.5\%) & 23 (50.0\%) \\
Performers              & 8 (36.4\%)  & 21 (45.7\%) \\
Conservatory students    & 6 (27.3\%)  & 6 (13.0\%) \\
\hline
\end{tabular}
\end{table}

Table~\ref{tab:demographics} shows that both rounds attracted evaluators with strong domain expertise, with particularly high representation from researchers, music technologists, and active performers. The final round exhibited a higher proportion of expert profiles overall, reflecting the more specialised audience present at the live session, while the preliminary round maintained broader participation.

\subsection{Live Contest Evaluation}

From the live contest audience of ISMIR 2025 attendees, 48 people responded as evaluators, providing real-time assessment during the contest. Each system rendered the unseen piece, with audience voting conducted during the concert on the evaluation form which was distributed prior to the first performance. The submission window closed 5 minutes after the last performance. The voting results were computed directly after the submission window closed, and announced to the audience.

\section{Participants and Models}

Nine systems from international research groups were submitted to RenCon 2025, representing diverse technical approaches. The participating systems come from 6 distinct countries (based on the institutional address of the first author) and span 23 years (based on the first publication date of the submitted version of the system).

\begin{table}[h!]
\centering
\caption{Participating Systems}
\label{tab:systems}
\resizebox{\columnwidth}{!}{\begin{tabular}{@{}lll@{}}
\toprule
\textbf{System} & \textbf{Year} & \textbf{Main Approach} \\
\midrule
VirtuosoNet & $2019^{3}$ & Hierarchical GRU with cVAE  \\
DirectorMusices & 2002\tablefootnote{The participating model is slightly improved from the published version.} & Rules + SVR-based Dynamics \\
Midihum & 2023 & XGBoost with 400 Features \\
Contin-U & 2025 & Image-to-Audio Transformer \\
ScorePerLockNAR & 2025 & Non-AR Transformer + Templates \\
RenConnoisseur & 2025 & Phrase/Contour Heuristics  \\
ElegantAIPianist & 2025 & Style-Adaptive Layer Norm \\
YQX+ & 2025 & Probabilistic Flow Matching \\
CueFreeExpressPedal & 2024 & Ensemble of Transformer Encoders \\
\bottomrule
\end{tabular}
}
\end{table}


The 2025 entries range from feature-engineered statistical models to transformer-based generative architectures. We categorize them roughly under the following four headings:

\textbf{Rule-Based and Statistical Learning}: The systems Director Musices and RenConnoisseur utilize expert-driven heuristics and pattern-based analysis of melodic contours and phrase structures. In Director Musices, the phrase arch rule was used for timing \cite{friberg1995matching} and a recent model using features and support vector regression for dynamics \cite{jones2023probing}. Midihum\footnote{\url{https://github.com/erwald/midihum}} bridges this approach with machine learning by using XGBoost's gradient-boosted trees trained on over 400 engineered features, primarily targeting velocity ``humanization'' while preserving the original score timing.

\textbf{Hierarchical and Probabilistic Models}:
VirtuosoNet (v1.1) \cite{Jeong2019VirtuosoNetPerformance} employs a hierarchical approach using Gated Recurrent Units (GRUs) and a conditional variational autoencoder (cVAE) to model expressive dependencies across note and phrase levels. Similarly, YQX+ \cite{zhouyqx} adopts a multi-scale probabilistic framework, introducing Conditional Flow Matching (CFM) to treat performance expression as a transport process from Gaussian noise to expressive deviations.

\textbf{Transformer-Based Architectures}: Modern sequence modeling is dominant in the 2025 entries. ElegantAIPianist \cite{chenautomated} uses a 4-layer bidirectional encoder and a 6-layer causal decoder with Style-Adaptive Layer Normalization (SALN) to inject composer-specific traits. ScorePerLockNAR \cite{zhai2025leveraging} differs by employing a non-autoregressive Transformer with a template-constrained pipeline to ensure 100\% structural fidelity to the score. CueFreeExpressPedal \cite{Worrall2024ComparativeMusic} utilizes an ensemble of five Transformer encoders to predict micro-timing and sustain pedal parameters from minimal input features.

\textbf{Cross-Modal Synthesis}:
Contin-U represents an alternative paradigm by bypassing intermediate symbolic MIDI representations. It employs a unified cross-modal Transformer that engraves MusicXML to images and leverages a direct image-to-audio synthesis pathway using residual vector quantization (RVQ) tokens.

\section{Results}

\subsection{Preliminary Round Rankings}

Table \ref{tab:prelim_results} shows the preliminary round results based on weighted average scores:

\begin{table}[htbp]
\centering
\caption{Preliminary Round Results}
\label{tab:prelim_results}
\begin{tabular}{@{}clc@{}}
\toprule
\textbf{Rank} & \textbf{System} & \textbf{Score} \\
\midrule
1 & DirectorMusices & 4.33/5.0 \\
2 & VirtuosoNet & 3.54/5.0 \\
3 & Midihum & 3.32/5.0 \\
4 & ElegantAIPianist & 3.19/5.0 \\
5 & Contin-U & 3.00/5.0 \\
6 & YQX+ & 2.83/5.0 \\
7 & ScorePerLockNAR & 2.53/5.0 \\
8 & RenConnoisseur & 2.53/5.0 \\
9 & CueFreeExpressPedal & 2.31/5.0 \\
\bottomrule
\end{tabular}
\end{table}

DirectorMusices achieved the highest preliminary score, demonstrating the continued effectiveness of well-crafted rule-based approaches. The neural systems showed strong but variable performance, with VirtuosoNet being ranked second overall.

\subsection{Live Contest Results}

Table \ref{tab:live_results} presents the live contest results, including the human performance for comparison. The model CueFreeExpressPedal was not able to participate in the final. 

\begin{table}[htbp]
\centering
\caption{Live Contest Results with Ranking Changes}
\label{tab:live_results}
\begin{tabular}{@{}clccc@{}}
\toprule
\textbf{Rank} & \textbf{System} & \textbf{Score} & \textbf{Prelim Rank} & \textbf{Change} \\
\midrule
1 & VirtuosoNet & 3.62/5.0 & 2 & $\uparrow$1 \\
2 & DirectorMusices & 3.06/5.0 & 1 & $\downarrow$1 \\
3 & Midihum & 2.90/5.0 & 3 & — \\
4 & Contin-U & 2.90/5.0 & 5 & $\uparrow$1 \\
5 & ScorePerLockNAR & 2.52/5.0 & 7 & $\uparrow$2 \\
6 & RenConnoisseur & 2.40/5.0 & 8 & $\uparrow$2 \\
7 & ElegantAIPianist & 2.08/5.0 & 4 & $\downarrow$3 \\
8 & YQX+ & 1.79/5.0 & 6 & $\downarrow$2 \\
\midrule
— & \textbf{Human} & \textbf{4.40/5.0} & — & — \\
\bottomrule
\end{tabular}
\end{table}

The results of the final round roughly agree with the ranking from the preliminary round, with VirtuosoNet, DirectorMusices, and Midihum being the top three models in each case. 
Among 48 responses, 36 (75\%) correctly identified the human performance, indicating that AI-generated expression has not yet reached the level of being indistinguishable from human performance. 

\begin{figure}[h]
  \centering
  \includegraphics[width=0.95\linewidth]{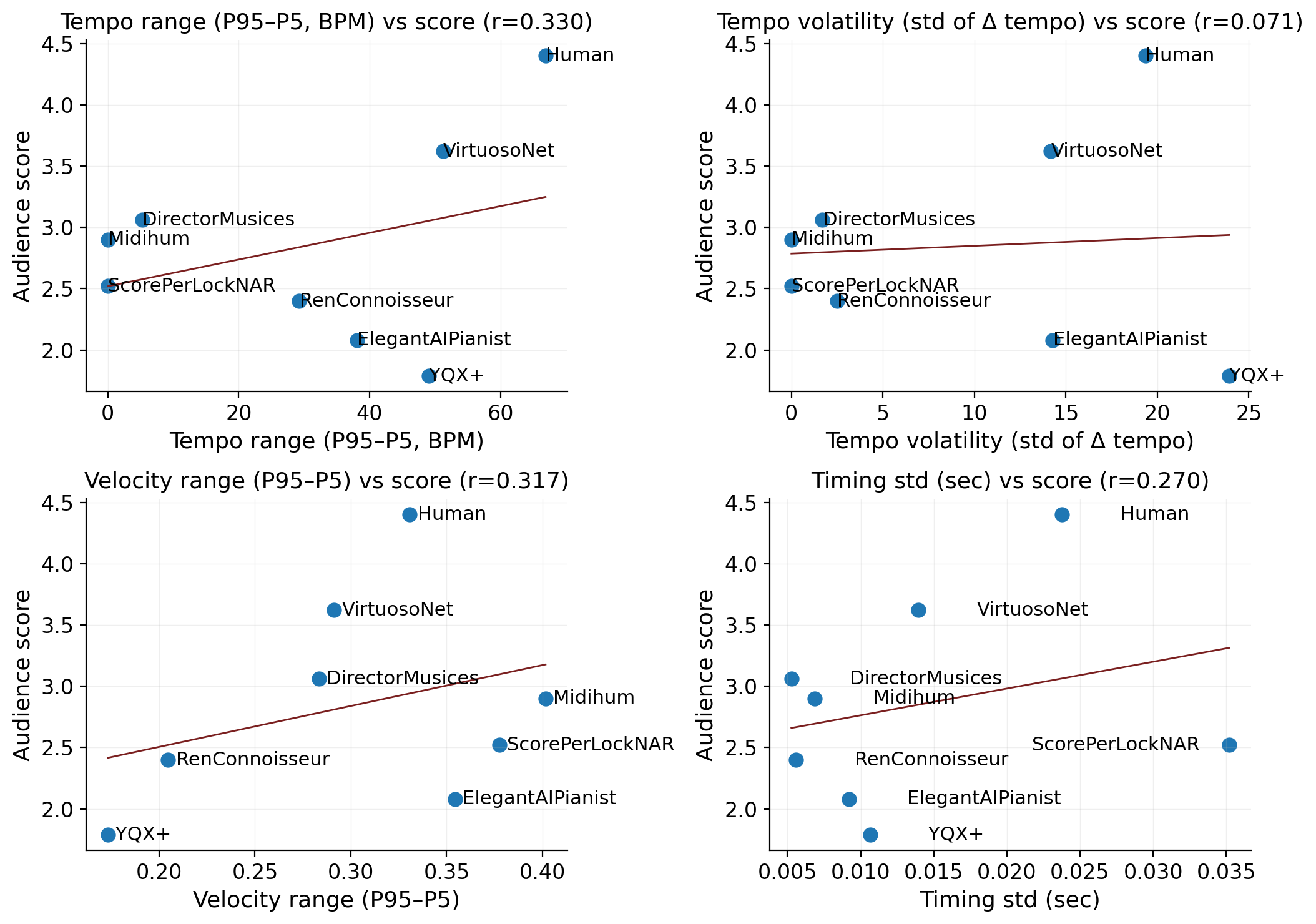}
  \caption{Each panel shows a summary metric (tempo, dynamics, timing, articulation) against audience score, with a linear trend and Pearson correlation $r$ reported in the title.}
  \label{fig:score-scatter}
\end{figure}

\begin{figure*}[h]
  \centering
  \includegraphics[width=0.95\linewidth]{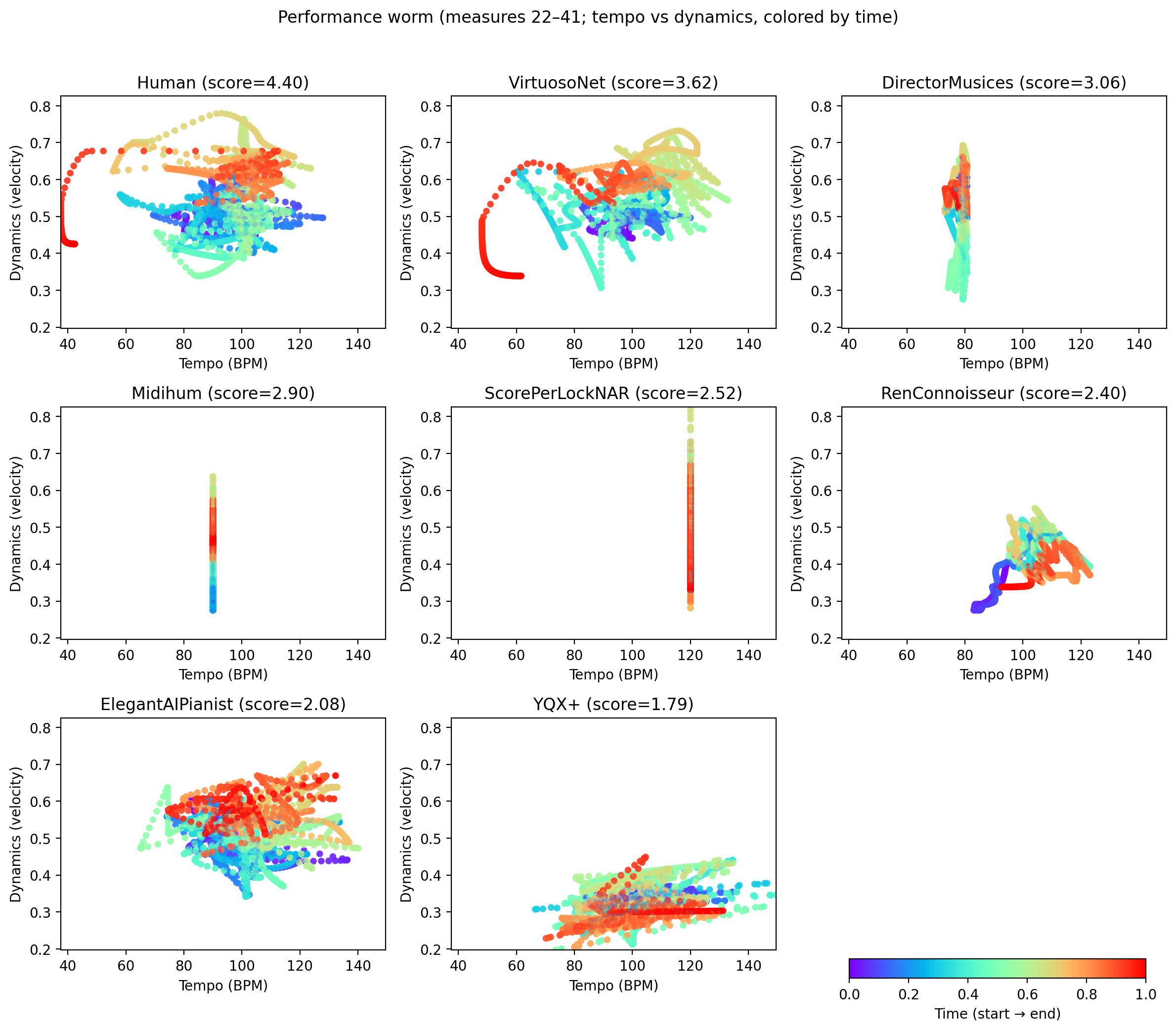}
  \caption{Each subplot shows tempo (x) versus dynamics (y), with color indicating progression from beginning to end of the second variation section. }
  \label{fig:performance-worm}
\end{figure*}

\section{Discussion and Reflection}

\subsection{Competition Design}

\subsubsection{Online audition}
As this was the first time an online audition phase was introduced, several challenges emerged, similar to those encountered when designing controlled listening tests.

One issue concerned \textbf{ordering}. To reduce potential bias, submissions were presented in a floating interface that avoided a fixed listening sequence. However, feedback from the evaluation questionnaire indicated that some evaluators found the floating windows difficult to grab and keep track of.

Another constraint was related to \textbf{anonymity}. Submissions were anonymised so that evaluators judged solely on performance quality instead of their familiarity with the work (most submissions are published systems). 
Technical reports were not displayed alongside submissions, since methodological details could reveal participant identities. As a result, the evaluator focussed on rendered outputs, and they were unable to rate the technical innovation.


\subsubsection{Live event}
Due to limited time during the live event, teams could not perform direct per-team adjustment at the venue. Although practical, this organizer-side intervention should be formalized and shared in advance in future editions.

We also note an operational issue during the event: MIDI from DirectorMuscies was played twice because the first playback level was too low. This incident further indicates that robust playback procedures and level-check protocols are needed.

For future editions, it would be beneficial to provide a pre-contest reference playback profile (or a sampled instrument proxy) so teams can audition their systems under conditions closer to the final instrument response before submission. More broadly, this points to an open problem for generative rendering systems: there is still no standard target or protocol for audio-level adjustment against real acoustic playback conditions.


\subsection{Evaluation Insights}

Evaluation of expressive performance has long been of research interest \cite{Morsi2024SimulatingLearning, Zhang2024HowDataset}. To analyze the use of expressive devices, we aligned each system’s MIDI performance to the reference score using \texttt{Parangonar}'s \texttt{DualDTWAlignment}\cite{Peter2024TheGlueNoteAlignment}, then extracted expressive performance parameters (tempo, dynamics or velocity, timing, and articulation) with \texttt{partitura} \cite{Grachten2019PartituraData} from the score-performance alignments. We excluded the audio-based Contin-U from analysis: After attempting automatic transcription, the resulting MIDI data led to unreliable alignment and distorted tempo results.

\paragraph{Expressive metrics vs. audience score (Fig.~\ref{fig:score-scatter}).}
Across the scatter grid, dynamics-related measures show the clearest positive association with audience score: broader velocity spread and higher velocity standard deviation tend to correspond to higher ratings. 
By contrast, tempo range and tempo volatility exhibit weaker or inconsistent relationships, indicating that large tempo excursions alone are not sufficient for higher perceived quality. Timing variability also provides a positive correlation.

\paragraph{Performance worm (Fig.~\ref{fig:performance-worm}).}
The performance worm \cite{Dixon2002PerformanceAnimation} representation reveals each system's expressive trajectory as a continuous shape in the tempo--dynamics plane. In Figure~\ref{fig:performance-worm} we have selected the region of the second variation (m.22-m.41), and the trajectory is built via PCHIP (monotone cubic) interpolator between each pair of onsets. High-scoring performances (notably Human and VirtuosoNet) trace coherent arcs with similarities in dynamics at comparable tempo regions, suggesting stable, intentional expressive planning rather than abrupt, erratic shifts. 
Lower-scoring systems show more fragmented clouds with less directional flow.
A notable finding is that Midihum and ScorePerLockNAR, which largely preserve score timing with minimal expressive rubato, still receive mid‑tier rankings. This suggests that a stable tempo is acceptable, or in fact preferable to poorly chosen tempo or timing deviations.

\section{Future Directions}

We would like to continue the competition and strengthen RenCon community building in the future.  Collaborating with the MIREX initiative allows us to explore new research directions beyond standard rendering. For example, we could introduce tasks like predicting the technical difficulty of a piece from its score, which has applications in both music education and automated curation.

Another plan for the future is moving toward more diverse data modalities. While most current systems focus on MIDI or audio, we want to include embodied performance data, such as the physical key and pedal movements captured by Yamaha Disklavier systems. The recent success of models like Contin-U shows that we can now work directly with score images to produce audio, effectively merging optical music recognition with expressive performance. This suggests that the competition should broaden its scope to include systems that process various inputs --- from scanned sheet music to physical performance captures.



\section{Conclusion}

RenCon 2025 marked the return of the expressive performance rendering competition and provided a contemporary snapshot of current approaches to score-to-performance rendering. The event brought together nine systems from six countries and engaged both online and live evaluators. The two-phase format enabled evaluation under both controlled online listening and concert-like conditions.

The results provide several empirical observations. First, across both rounds, rule-based and learning-based systems remained competitive, with different models excelling under different evaluation settings. Second, the human reference performance received the highest rating (4.40/5.0), and 75\% of evaluators correctly identified it, indicating that current systems are still perceptibly distinguishable from human performance. Third, analysis of aligned MIDI data showed that dynamics-related variability correlated more consistently with audience ratings than large tempo deviations, suggesting that dynamic shaping may play a stronger perceptual role than extreme rubato in this evaluation context.

RenCon 2025 provides an updated reference point for how expressive rendering systems are currently evaluated and perceived. The competition format, collected evaluation data, and documented procedures offer a basis for refining future assessments and for studying how listeners perceive expressive performance generated by computational systems.

\section{Acknowledgments}

We thank the ISMIR 2025 organizing committee and MIREX coordinators for supporting RenCon's revival. We are also grateful to previous RenCon organisers Katayose Haruhiro, Mitsuyo Hashida and Robert Bresin, for initiating and continuing to support the cause. We thank Kenzi Noike for their suggestions and feedback throughout the competition. We thank Patricia Hu for creating audio recordings from the preliminary round MIDI files using a Disklavier, as well as the volunteer Yuncong Xie for assisting with the live event. Special thanks to all participating teams, evaluators, and audience members who made this competition possible. 
This work was supported by funding from UK Research and Innovation [grant number EP/S022694/1].

\section{Ethical Standards}

RenCon 2025 involved voluntary participation from system developers and audience evaluators in the context of an academic competition and public concert. No personal identifying data was collected beyond optional, high-level demographic descriptors (e.g., musical background), and all evaluation responses were analysed in aggregate. Participants were informed of the evaluation procedure and consented to their contributions being used for research and reporting purposes.

The competition design and evaluation protocol followed established practices from previous RenCon, ISMIR, and MIREX benchmarking activities. No experiments involving vulnerable populations, medical intervention, or animal subjects were conducted. The authors declare no competing financial or non-financial interests.




\bibliographystyle{ACM-Reference-Format}
\bibliography{sample-references, huan_ref}

\appendix

\end{document}